A NOVEL DELAY-TIME ENLARGED 3-D GRAVITATIONAL WAVE DETECTION SYSTEM

Andrew Huang

Maranatha High School, Pasadena, CA 91105

Email: AndrewHuang2014@gmail.com

ABSTRACT

A novel delay-time enlarged 3-dimensional gravitational wave (GW) detection system is presented. The operation principle is described. The basic specification requirements for all the critical components are analyzed. The whole system consists of three optical fibers along three axes, a narrow linewidth ultra-stable laser, an ultra-stable radio frequency (RF) source, three recirculating optical fiber loops, three phase monitoring/stabilizing units, three phase detectors, and a computer based data analysis unit. With the given specifications of the critical components, the whole system may achieve $10^{-22}$ phase sensitivity, and therefore can be used for the GW detection. This is the first time, to the best of our knowledge, one has implemented an optical fiber as the delay-time enlarged transmission medium for a self-delayed interferometer and the first 3-dimensional self-delayed interferometer. Since optical fiber is used as the transmission medium and the recirculating optical fiber loop for increasing the phase sensitivity, the whole system can be built both compactly and cost efficiently, allowing a 3-dimensional self-delayed interferometer for GW detection to be created.

## I. INTRODUCTION

The detection of the gravitational wave (GW) is one of the essential topics for the research of the universe's history, cosmology, and astrophysics, as well as fundamental physics laws. Over 100 years have passed since Albert Einstein predicted the existence of GW in his general theory of relativity [1], [2]. Many individuals (including Joseph Weber [3], [4], M.E. Gertsenshtein [5], Rainer Weiss [6], Ronald Drever [7], and Kip Thorne [8], et al) and groups/projects (including LIGO [9], LISA [10], TAMA300 [11], Virgo [12], AIGO [13], GEO600 [14], and Swirl Antennas detection system [15] etc.) have been working on the search for the GW in past century. Most mentioned measurement systems are based on a Michelson interferometer (except [10] and [15]), which use a laser as the signal source. The laser light is split into two beams and passed through two perpendicular high vacuum arms before being recombined. Once a GW passes through the measurement system, the space-time field will be changed (based on Einstein's general theory of relativity), which will then cause a change of the length difference between the two arms. As such, the phase difference between two laser beams will change, allowing us to receive and analyze the GW's information and data.

Since the distance of the GW source is too far from the earth (in $10^9$ lightyear order), the GWs are extremely weak when they reach the earth – only causing a change in space on the order of one part in $10^{22}$ (i.e. the relative space length change within $10^{-22}$ order), they have never been directly detected until late 2015. A US-led international team built the Laser Interferometer Gravitational-Wave Observatory (LIGO), then updated the original LIGO to Advanced LIGO [16]. This project took more than 20 years, with costs exceeding $620 million in US dollars [17], before the GWs were ever detected

(specifically on September 14, 2015 [18] and December 26, 2015 [19]). A space-based GW detector (like LISA) will cost even more than $1 billion in US dollars [20]. Despite LIGO including only two interferometers, for real space GW detection, we need 3-dimensional information, which means we have to build more interferometers (costing over $300 million in US dollars per interferometer).

A more cost efficient 3-dimensional GW detection system and method is needed for GW detection.

In this paper, a novel delay-time enlarged 3-dimensional GW detection system is presented. The operation principle is described (in Section II). The basic theoretical analysis of the system is given, and the specification requirements for all the critical components are proposed (both in Section III). This is the first time, to the best of our knowledge, that an optical fiber has been used as the delay-time enlarged transmission medium for a self-delayed interferometer and the first 3-dimensional self-delayed interferometer created. Since optical fiber is used as the transmission medium and the recirculating optical fiber loop for increasing the phase sensitivity, the whole system can be built both compactly and cost efficiently, allowing a 3-dimensional self-delayed interferometer for GW detection to be created.

## II. SYSTEM SETUP AND OPERATION PRINCIPLE

The whole 3-D GW detection system consists of three optical fibers along three axes (X, Y, and Z-axis), a narrow linewidth ultra-stable laser, an ultra-stable RF source, three recirculating optical fiber loops, three phase monitoring/stabilizing units, three phase detectors, a computer based data analysis unit, and related support units. The three optical fibers along the three axes are used for picking up the space/length change caused by GW. The narrow linewidth ultra-stable laser is used for carrying the ultra-stable RF signal. The ultra-stable RF source is used for detecting the length change (by detecting its phase change that will accompany the length change) caused by GW. The recirculating optical fiber loop is used for enlarging the equivalent delay length and time (therefore increasing the system phase sensitivity). The phase monitoring/stabilizing unit is used for stabilizing the recirculating fiber loop length. The three phase detectors are used for detecting the phase change from three axes. The analysis unit is used for analyzing and synthesizing the data from the phase detectors and generating the 3-D data of the GW. The functions of each unit are described as following:

The block diagram of the delay-time enlarged 3-dimensional GW detection system is shown in Fig. 1. The three branches (Branch X, Y, and Z) have been designed with the exact same structure, and are perpendicular to each other. They run along the X, Y, and Z axes so the direction of an incoming GW may be identified, regardless of the orientation. In Fig. 2, a narrow linewidth laser is used as the low noise and ultra-stable signal source carrier, which is modulated by an RF signal source through an electronic-optical modulator (EOM), so an ultra-stable synchronized signal ($S_X$, $S_Y$, and $S_Z$) can be generated and distributed to the three axes (X, Y, and Z) for GW detecting.

In Branch X (refer to Fig. 3; although all three branches are exactly same, only Branch X and its details are described), $S_X$ will pass through Arm X (a segment of optical fiber) first. If a GW passes through Arm X, the length of the fiber will be changed, so the transmission time and phase of $S_X$ will be changed as well. Then $S_X$ is fed into Loop X for recirculating for a time period $\Delta T$. Meanwhile the $S_X$ is continuously launched into Arm X. After time period $\Delta T$, the delay time and phase difference are amplified, so we may see the system phase sensitivity increase. At this time, both the new signal in Arm X and the delayed original signal in Loop X are sent to the Phase Detector for phase comparison (also known as phase discrimination). The enhanced phase change signal can then be compared with that of the other branches (Branch Y and Branch Z), so we can discern the GW's information (direction, amplitude, frequency, occurrence time, etc.) through the Data Analyzer.

Arm X consists of a segment of optical fiber with a typical length of 100 km. Since it will be wound up on a fixture that is 10 meters long, it can be easily mounted vertically (for Arm Z setup). As such, we may build a 3-dimensional detection system.

Fig. 4 shows the details of the recirculating optical fiber Loop X. The loop switches $SW_{L1X}$ and $SW_{L2X}$ are used for receiving and recirculating the original signal $S_X$. Once the Loop switch controller (Loop SW Controller) turns $SW_{L1X}$ into position "A" and $SW_{L2X}$ to position "B", the Loop X may receive $S_X$. If $SW_{L1X}$ is in position "B" and $SW_{L2X}$ is in position "B", the $S_X$ will recirculate inside the Loop X. The recirculating turns (therefore corresponded delayed time) may be controlled by Loop SW controller. Once $SW_{L1X}$ is in position "B" and $SW_{L2X}$ is in position "A", the $S_X$ will be sent out of the loop for phase comparison and GW info detection. An optical amplifiers (OA) is used for compensating the insertion loss caused by the optical fiber. A 3-dB coupler OC3 is used for sending Sx into the loop and coupling the Sx out of the loop.

The Loop Length Stabilizer, optical couplers OC1 and OC2, and the Loop Length Compensator are used for stabilizing the loop length (refer to Fig. 4). The stabilizing signal Ss is sent into the loop from OC1 with a different wavelength from Sx. After traveling one loop cycle, Ss is tapped down from OC2 and compared with the original Ss at the loop length stabilizer. The loop length change info will be detected and a control signal will be sent to the loop length compensator to compensate the length change and keep whole loop as a fixed length.

The whole system runs on a single frequency signal, so the optical wavelength can work at a zero dispersion area (for instance, use 1310 nm wavelength for the regular single mode fiber). The chromatic dispersion and the polarization dispersion will not be the critical problem. If any dispersion becomes the main noise source for the system detection, we can use the regular compensation technology (typically, using a high negative dispersion fiber module to compensate the positive dispersion caused by the loop fiber) to eliminate it.

Since each loop can only take a piece of signal $S_X$ (about the length of Arm X), we need multiple loops to carry different points of the signal $S_X$. So a more functional GW detection system should be built with multiple channels to pick up signals of different time periods (shown in Fig. 5), which allows the system to collect all of the 3-D GW information.

III.  THEORETICAL ANALYSIS

Since on the Earth's surface (ground), we can only pick up GWs within the frequency range from around 10 Hz to 500 Hz, we may take 100 Hz as the typical frequency ($f_{GW}$ = 100 Hz) in our theoretical analysis below. The largest challenge for GW detection is the fact that GWs are extremely weak when they reach the earth – only causing a change in space/distance of order of magnitude 1 part in $10^{22}$ (i.e.

the relative space length change within $10^{-22}$ order). Our analysis will take these numbers (frequency of 100 Hz and magnitude change within $10^{-22}$ order) as the reference and determine the operation conditions (required specification) of each key unit in our system.

### 3.1 Signal Source

Let *s(t)* be the original RF reference source signal

$$s(t) = S_o Sin(2\pi f t + \phi_o + \Delta\phi_s)$$

After modulating a laser with frequency $f_{LD}$ and passing through a fiber with length $L_A$, we have

$$S_L(t, L_A) = K S_o Sin\left(2\pi \frac{L_A}{\lambda} + 2\pi f t + 2\pi f_{LD} t + \phi_o + \Delta\phi_s\right) \quad (1)$$

where $L_A$ is the length of each arm of the interferometer, $\phi_o$ is the original phase, and $\Delta\phi_s$ is the phase fluctuation (the phase noise of the source signal). We may set $\phi_o$ as 0 through the calibration of all branches (axes). For an ideal signal source, $\Delta\phi = 0$. If we use the intensity modulation (IM) method through an external electronic-optical modulator (EOM) and a narrow linewidth laser, the phase noise contribution from the laser can be ignored. So we can also ignore the modulating / demodulating procedure; therefore we have

$$S_L(t, L_A) = S_o Sin\left(2\pi \frac{L_A}{\lambda} + 2\pi f t\right) \quad (2)$$

For an optical wave modulated by *s(t)*, travelling through a single mode fiber (SMF) with index of refraction *n*, we have the transmission speed

$$v = \frac{c}{n} = f\lambda \quad (3)$$

and

$$S_L(t, L_A) = S_o Sin\left(2\pi \frac{n}{c} f L_A + 2\pi f t\right) \quad (4)$$

Once the GW makes the optical fiber length change by $\Delta L$, *s(t)* will become

$$S_L(t, \Delta L) = K S_o Sin\left[2\pi \frac{n}{c} f (L_A + \Delta L) + 2\pi f t\right]$$

$$S_L(t, \Delta L) = K S_o Sin\left[2\pi f \frac{n}{c} L_A \left(1 + \frac{\Delta L}{L_A}\right) + 2\pi f t\right] \quad (5)$$

Let

$$\Delta\phi = 2\pi \frac{n}{c} f \Delta L \ (rad.) \ = 360 \frac{n}{c} f \frac{\Delta L}{L_A} L_A \ (°) \quad (6)$$

The typical distance change ratio *k* caused by a GW will be

$$k = \frac{\Delta L}{L_A} = \frac{\Delta \tau}{\tau} = 10^{-2} \tag{7}$$

where $c$ is the speed of light ($3 \times 10^8$ m), $n$ is the index of refraction (1.48 typically for SMF). Let the modulating frequency $f$ be 100 GHz ($10^{11}$ Hz), and we have

$$L_A = \Delta\phi \Big/ \left(360 \frac{n}{c} f \frac{\Delta L}{L_A}\right) = \Delta\phi \Big/ \left(360 \cdot \frac{1.48}{3 \cdot 10^8} \cdot 10^{11} \cdot 10^{-22}\right)$$

$$= 5.63 \cdot 10^{13} \cdot \Delta\phi \quad (km) \tag{8}$$

### 3.2  EOM and Photodetector

For the EOM modulation performance – called "S" curve, which is usually about 10 V per 180 ° (i.e. $V_\pi$ = 10 V). We may increase the amplitude of the modulation signal by 10 times but limit the total amplitude as unchanged (make it very much like a square wave). This way, the slope of the cross-zero point will increase 10-fold (see Fig. 6). The equivalent input signal (vs. the phase noise) fed into the EOM will increase tenfold, so we have $V_{\pi eq}$ = 100 V. Let the resolution of the photodetector (PD) and the following circuit be 0.1 µV (i.e. it can identify a $10^{-7}$ V signal), therefore we have

$$\Delta\phi = \frac{\Delta V}{V_{\pi eq}} \pi = \frac{0.1 \, \mu V}{100 \, V} 180° = 1.8 \cdot 10^{-7} \, (°) \tag{9}$$

and

$$L_A = 5.63 \cdot 10^{13} \cdot 1.8 \cdot 10^{-7} = 10^7 \quad (km) \tag{10}$$

which means that, if we use a $10^7$ km optical fiber as the delay line, the system can identify a length change with a $10^{-22}$ order (caused by GW)!

### 3.3  How to Make a Long Delay Line

To make a $10^7$ km SMF delay line, we may use a recirculating optical fiber loop (shown in Fig. 4). If the loop has total length of 1000 km (10 segments cascaded, each segment including 100 km SMF and one optical amplifier), we can let the signal travel inside the loop for $10^4$ turns and define it as one recirculating cycle. So the original signal will run a total of $10^7$ km before it is sent out to compare with the new signal. Therefore the recirculating optical fiber loop may be equivalent to a $10^7$ km long delay line [21]. Once the GW passes through and the phase difference between the delayed signal and new signal has been enhanced, the system may detect and display the information of the phase change.

### 3.4  The Phase Noise of the RF Source

An ultra-low noise RF source is essential for GW detection. Since we take the phase change information to detect the GW, the phase noise of the RF source will directly affect the measurement and must be lower than the phase change caused by GW. A typical GW has a frequency of around 100 Hz and a $10^{-22}$ relative amplitude change. So the phase noise of the RF source has to be lower than $10^{-22}$ at 100 Hz off the modulation frequency. If we take an oven-controlled crystal oscillator (OCXO) as the RF signal source, its typical specifications will be: operating frequency 10 MHz, phase noise -160 dBc/Hz at 100 Hz (off frequency). The phase noise may be improved (by dividing down the frequency) or degraded (by multiplexing the oscillation frequency). The relationship of the phase noise change between a multiplexed frequency and its original oscillation frequency may be expressed as

$$\Delta N_P = 20\, Log_{10}\, (N) \quad (dB) \tag{11}$$

So if the frequency doubles (i.e. $N = 2$), the phase noise of output signal (with the frequency 2 times higher than the original signal frequency) will be 6 dB higher (worse) than the original signal. Similarly, for the OCXO operating at 10 MHz and with phase noise of -160 dBc/Hz at 100 Hz, the phase noise will drop down to -240 dBc/Hz at 100 Hz if the operating frequency is reduced to 1 kHz (i.e. $N = 1/10000$). -240 dBc/Hz means the phase noise will be lower than $10^{-24}$ in ratio. So we may use such an OCXO oscillator as our RF source.

### 3.5 Considerations of Thermal Fluctuation (Thermal Noise) of Optical Fibers

Since both the transmission arm and the recirculating loop use optical fibers as the transmission medium, the fiber stability (particularly the thermal stability) will be a critical factor for the system's sensitivity and stability. Following are several considerations and approaches to stabilizing the optical fiber arm and loop, and ensuring the system sensitivity will be high enough to detect the gravitational wave:

#### 3.5.1 Phase Noise, Thermal Noise and Carrier Signal Frequency

According to the definition of the phase noise [22]:

$$S_\phi(f) = \frac{\phi^2(f)}{BW} \tag{12}$$

where $\phi(f)$ is the instantaneous phase departure from a nominal phase. For a different operating frequency $\nu_0$ (with different cycle $T$), the same time fluctuation $\Delta t$ will cause a different phase noise. The relationship will be similar to equation (11).

The thermal noise caused by optical fiber phase noise will typically be -125 dB at 100 Hz for a 1310 nm or 1550 nm (laser) operating wavelength [23]. The frequency of a 1310 nm light wave traveling on a standard single mode fiber (SMF) with refractive index 1.457 will be:

$$f_o = \frac{c}{n\lambda} = \frac{3 \cdot 10^8}{1.457 \cdot 1310 \cdot 10^{-9}} = 1.57 \cdot 10^{14} \quad (Hz) \tag{13}$$

or 157 THz. Its duty cycle T will be

$$T = \frac{1}{f} = \frac{1}{1.57 \cdot 10^{14}} = 6.4 \cdot 10^{-15} \ (second) \quad (14)$$

or 6.4 *fs*. We can see that any small time delay fluctuation will cause a huge phase noise. For instance, a 6.4 x $10^{-17}$ second time delay variation will cause 3.6° phase uncertainty.

In our GW detection system, instead of light waves, we take RF waves as the GW detecting probe. The RF frequency will be much lower than the light wave frequency. If taking a 1 GHz RF signal as the detecting source, we have

$$f_{RF} = 10^9 \ Hz = \frac{10^9}{1.57 \cdot 10^{14}} f_o = 6.4 \cdot 10^{-6} \cdot f_o \quad (15)$$

which means that (based on equation 12); if the fiber thermal noise that is causing phase noise for an optical signal (1310 nm) at -125 dB [23], it will reduce to

$$S_{\phi RF}(f) = -125 + 10 \ Log \ (\frac{f_{RF}}{f_o})^2 \quad (16)$$

$$= -125 + (-120) + 16$$

$$= -229 \ (dB \ r/Hz) \quad (17)$$

Therefore, if the power spectrum $S_\phi(f)$ of the phase change caused by GW is higher than $10^{-22.9}$ at 100 Hz off the carrier frequency, the detector at the receiver end can recognize the GW signal.

### 3.5.2 Thermal Noise on Phase Modulation System

Kjell Blotekjaer has systemically analyzed the influence of optical fiber thermal noise on long-distance coherent communication systems [24]. The influence of the thermal noise on GW detection systems will be similar to that on a DPSK system. From the theoretical analysis and typical number calculation, he concluded that the "thermal noise is not a problem with fiber lengths less than 10 000 km. Hence, thermal noise is unlikely to be of importance for DPSK coherent communication systems".

### 3.5.3 Improvement of the Thermal Performance of the Optical Fiber

The typical thermal extension coefficient of a standard single mode fiber (SMF) is 7 ppm per °C (i.e. 7x$10^{-6}$ / °C). So even if we stabilize the fiber environmental temperature within 1 m°C (i.e. ΔT < $10^{-3}$ °C), the fiber stability may get 7x$10^{-9}$, which is still not enough for detecting the phase / length change information with a $10^{-22}$ order. Fortunately the thermal change is a slow process (compared to the ~100 Hz change). Also, if we take specialized low thermal coefficient of delay (TCD) fiber, like a Sumitomo's product, then this problem will be eliminated. Because the TCD slope of Sumitomo's fiber will change from negative to positive at about 5 °C [25], which means there is a point (around 5 °C) at which the fiber TCD will be exact zero. So we may control the environmental temperature at an optimized point (~ 5 °C) and make the optical fiber operate at the zero TCD condition. Therefore the fiber's thermal noise may be suppressed to a negligible level.

### 3.5.4 The Improvement of the System Signal-to-Noise Ratio (SNR)

Since the thermal noise is a type of white noise. We may add on a band-pass filter (BPF) to receive signals in the GW frequency range (50 ~ 150 Hz) and improve the signal-to-noise ratio (SNR) of whole detection system.

### 3.5.5 Closed loop monitoring / compensating

Because the length of time in which the light waves are travelling in the recirculating loop is much longer than the optical arm, the loop's thermal noise will dominate the whole system's thermal stability. Using WDM (Wavelength Division Multiplexing) technology, we can implement a calibration channel with a different wavelength (from the GW detecting wavelength) to continuously monitor the loop length change, which including a compensation unit to keep the whole loop at a constant length even when temperature changes.

With the considerations of the five approaches above, the influence of optical fiber thermal noise on the system will be eliminated.

## 3.6 Fiber Length and Multiple Channels

Signal sampling and combining: assuming the optical fiber length of each arm (i.e. Arm X, Y, and Z) is 100 km and the GW frequency is 100 Hz (see Fig. 7), each arm can pick up only one twentieth of the whole cycle of the GW signal. So we can use 4 ~ 10 channels on each axis/arm to recover the GW shape.

The relationship between the arm timing and the loop timing is shown in Fig. 8. For increasing the delay time (to improve the phase sensitivity) by using the recirculating loop, we will lose some GW cycles. To reduce this loss, we may increase the length of the arm and/or the channel numbers, and/or take wavelength division multiplexing (WDM) technology to reduce the ratio of $T_L$ over $T_A$ so it may cover a larger time period.

## IV. CONCLUSIONS

The GW detection is one of the most important research fields for astrophysics, cosmology, and theoretical physics. The whole world has been searching the GW for 100 years since Albert Einstein predicted its existence. In this paper, a novel delay-time enhanced 3-dimensional GW detection system has been presented. The operation principle has been described. The basic specifications requirements for all the critical components have been analyzed. With the typical specifications of the critical components, the whole system can achieve $10^{-22}$ phase sensitivity, and therefore may be used for the GW detection. Since optical fiber is used as the transmission medium and the recirculating optical fiber loop for increasing the phase sensitivity, the whole system can be built both compactly (from ~ 4 km size reduced to ~ 10 m size) and cost efficiently (from ~ $300 million reduced to < $3 million), allowing a 3-dimensional self-delayed interferometer for GW detection to be created. This system can be also used for any other phase / length change monitoring and detection applications.

## V. ACKNOWLEDGEMENTS

The author likes to thank Prof. Yuk Yung of California Institute of Technology (Caltech), Prof. Sally Newman of Caltech, Prof. Yanbei Chen of Caltech, and Prof. Matthew Evans of Massachusetts Institute of Technology (MIT) for reviewing this paper and giving very helpful comments and suggestions. The author also likes to thank Dr. Shouhua Huang of NASA Jet Propulsion Lab (NASA-JPL) for his encouragement and help in writing this paper.

REFERENCES

1. A. Einstein, "Näherungsweise Integration der Feldgleichungen der Gravitation (Approximate integration of the field equations of gravitation)", Sitzungsber. K. Preuss. Akad. Wiss. Vol.1, pp. 688-696, 1916.
2. A. Einstein, "Über Gravitationswellen (On gravitational waves)", Sitzungsber. Preuss. Akad. Wiss., Vol. 1, pp. 154-167, 1918.
3. Joseph Weber, "Gravitational Radiation", Phys. Rev. Lett., Vol.18, pp.498-501, March 1967.
4. J. Weber, "Gravitational Wave Detector Events", Phys. Rev. Lett., Vol.20, pp.1307-1308, June 1968.
5. M.E. Gertsenshtein, V.I. Pustovoit, "On the Detection of Low Frequency Gravitational Waves", J. Exptl. Theoret. Phys., Vol.43, pp.605-607, August 1962.
6. R. Weiss, "Electromagnetically Coupled Broadband Gravitational Antenna", MIT RLE QPR, No.105, April 1972.
7. S. J. Augst, R. W. P. Drever, "Measurements of Mechanical Q in Levitated Paramagnetic Crystals", Third Amaldi Conference, Caltech, 2000.
8. K.S. Thorne, "*Black Holes and Time Warps: Einstein's Outrageous Legacy*", W.W. Norton & Company, New York, 1994.
9. D. Sigg, "Gravitational Waves", Proc. of Theoretical Advanced Study Institute (TASI) 1998.
10. G. D. Racca, et al, "The LISA Pathfinder Mission", Space Sci. Rev., vol.151, pp.159-181, 2010.
11. M. Ando, K. Tsubono, "TAMA Project: Design and Current Status", Proceedings of the Third Edoardo Amaldi Conference, 2000.
12. B. Caron, et al., "The Virgo interferometer", Class. Quantum Grav., 14, pp.1461–1469, 1997.
13. P. Barriga, et al, "AIGO: a southern hemisphere detector for the worldwide array of ground-based interferometric gravitational wave detectors", Class. Quantum Grav., 27, p.084005, 2010.
14. H. Lück, et al, "The GEO600 project", Class. Quantum Grav., 14, pp.1471–1476, 1997.

15. R. Courtland, "Swirly Antennas See the Ancient Cosmos", IEEE Spectrum, pp.9-11, April 2016.
16. J. Aasi, et al, "Advanced LIGO", Classical and Quantum Gravity, vol.32, 2015, 074001.
17. D. Castelvecchi, "Hunt for Cosmic Waves to Resume", Nature, Vol.525, pp.301-302, 17 September 2015.
18. B. P. Abbott, et al, "Observation of Gravitational Waves from a Binary Black Hole Merger", Phys. Rev. Lett. 116, 061102 (2016).
19. B. P. Abbott, et al, "GW151226: Observation of Gravitational Wave from a 22-Solar-Mass Binary Black Hole Coalescence", Phys. Rev. Lett. 116, 241103 (2016).
20. E. Gibney, "Space-based Detector Passes Teat Drive", Nature, vol.531, p.20, 3 March 2016.
21. S.-H. Huang, X.Y. Zou, E. Park, and A.E. Willner, "9,000-km WDM Transmission of Three 2.5-Gbit/s Channels Covering a 5-nm Wavelength Range", Topical Meeting on Optical Amplifiers and Their Applications, Proceedings, paper ThD14, Davos, Switzerland, June, 1995 (Optical Society of America, Wash., D.C., 1995).
22. John R. Vig, et al, "IEEE Standard Definitions of Physical Quantities for Fundamental Frequency and Time Metrology — Random Instabilities", IEEE Std 1139-1999, March 26, 1999.
23. R. E. Bartolo, et al, "Thermal Phase Noise Measurements in Optical Fiber Interferometers", IEEE J. Quantum Electron., vol.48, no.5, pp.720-727, May 2012.
24. Kjell Blotekjaer, "Thermal Noise in Optical Fibers and Its Influence on Long-Distance Coherent Communication Systems", J. *Lightwave Technol.,* vol. LT-10, no.1, pp. 36-41, 1992.
25. John W. Dreher, "Phase Stability of ATA Fiber Optic Cables", ATA Memo 55, March 19, 2003.

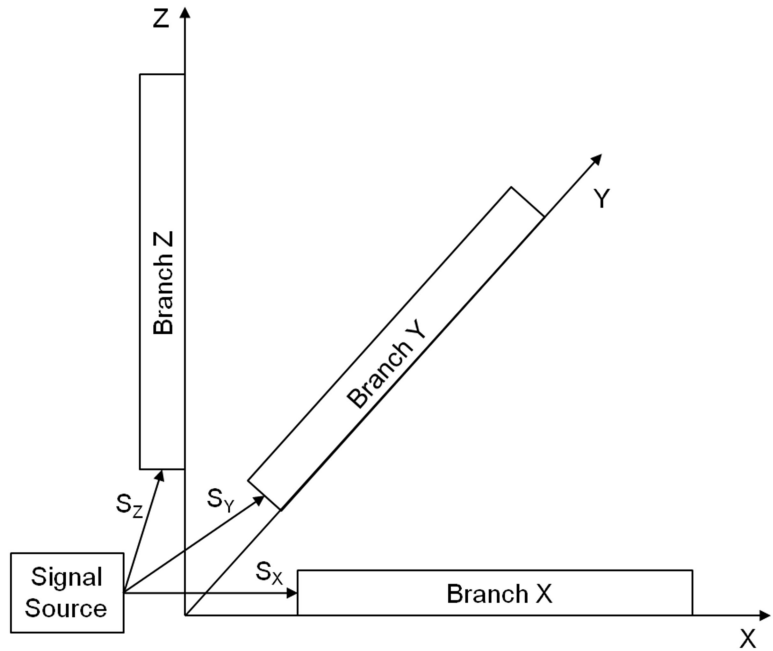

Fig. 1 Block diagram of the delay-time enlarged 3-dimensional GW detection system

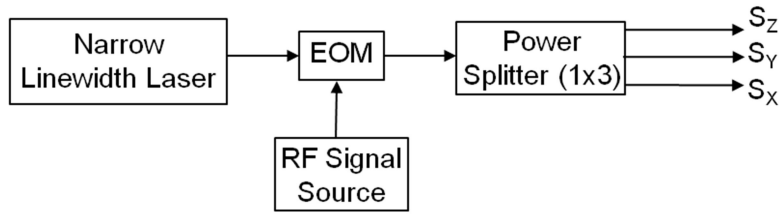

Fig. 2 Signal source of the 3-D GW detection system

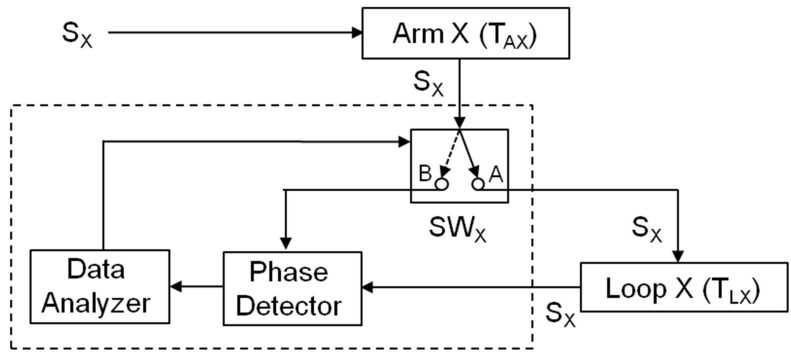

Fig. 3 One branch (Branch X) of the GW detection system

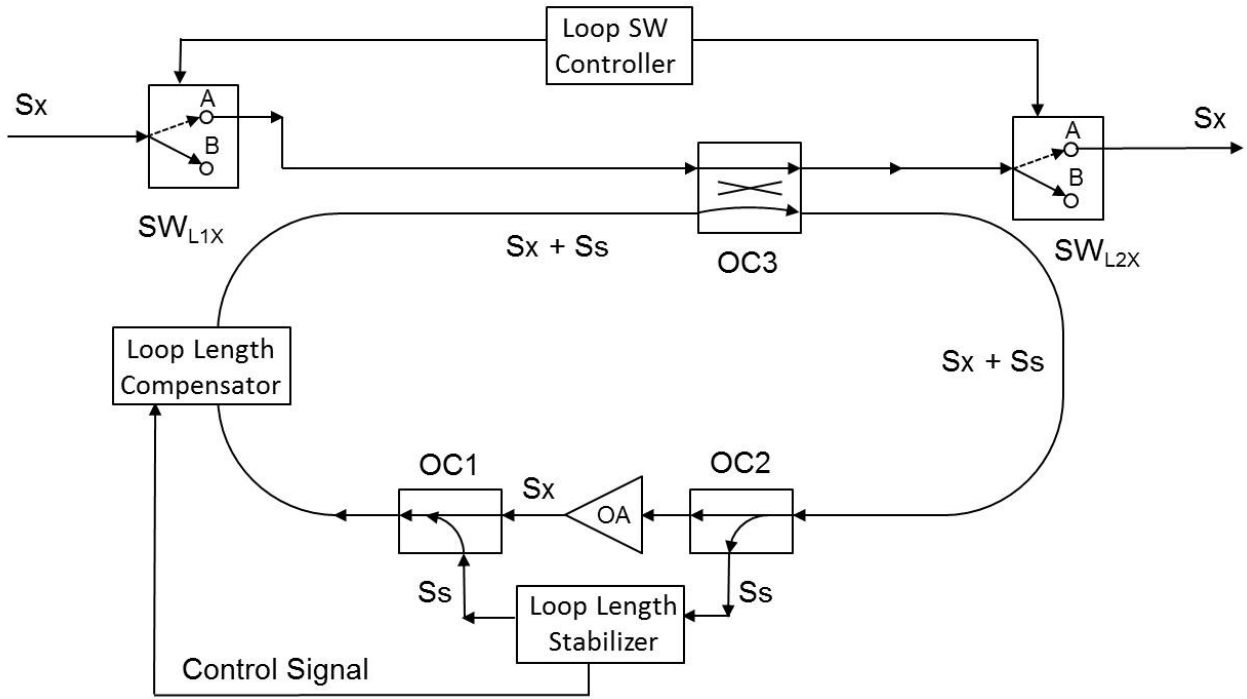

Fig. 4 The recirculating optical fiber loop (Loop X)

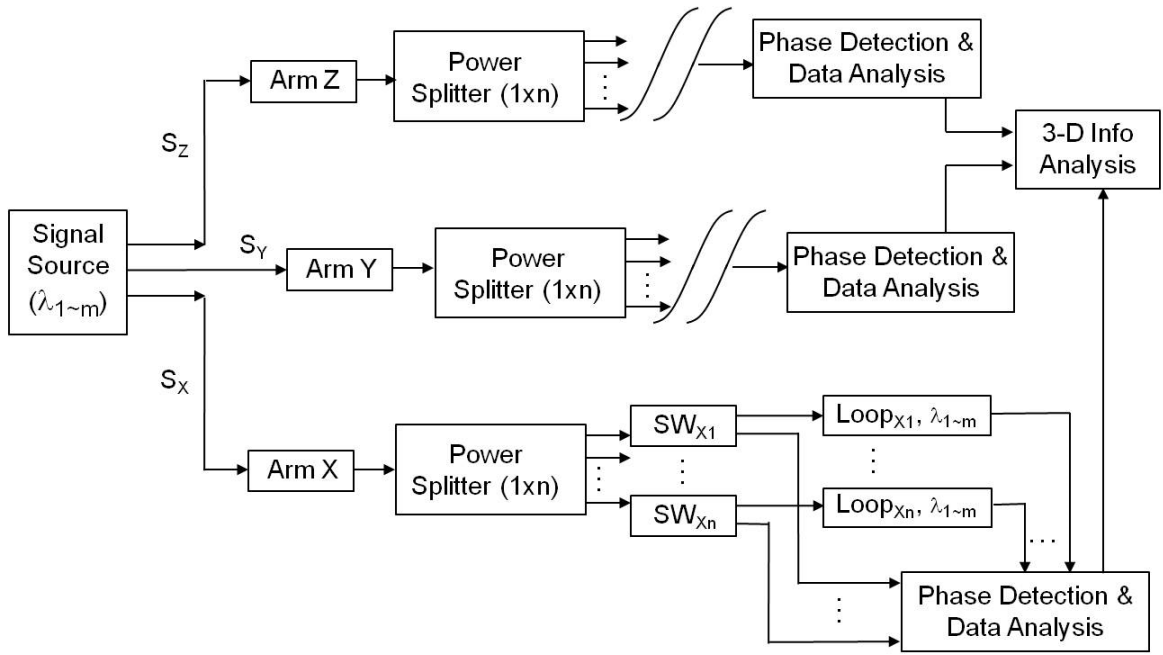

Fig. 5 Multiple channel 3-D GW detection system

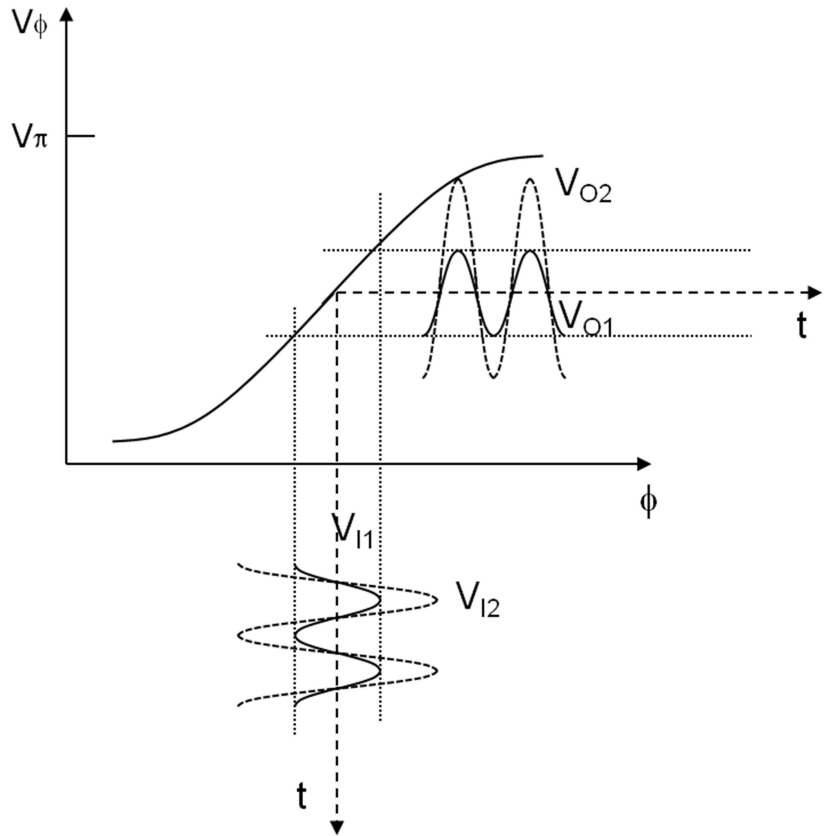

Fig. 6 Modulation (EOM) and Demodulation (Photodetector) operation curve

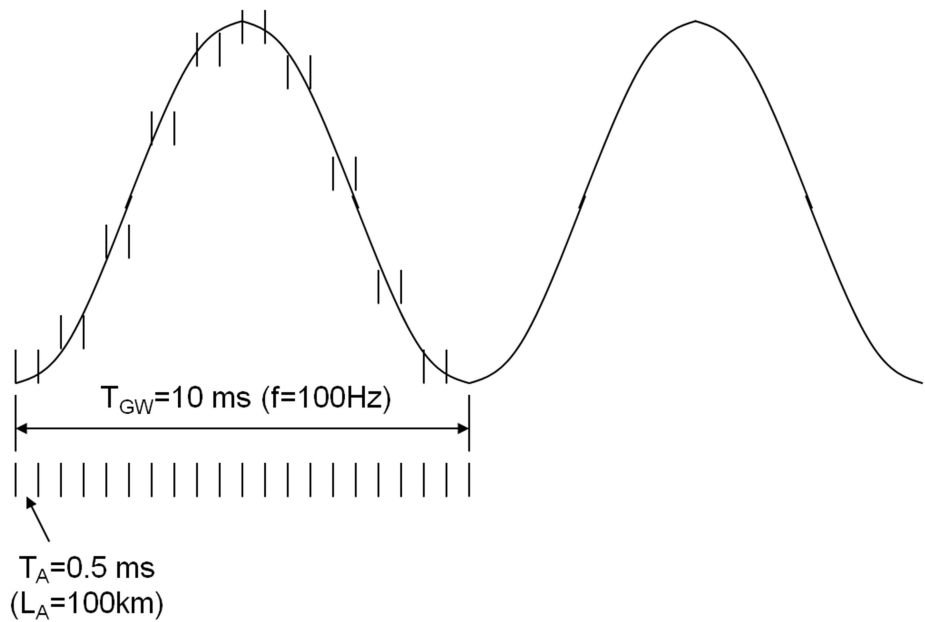

Fig. 7 Time relationship of the signal sampling and combining

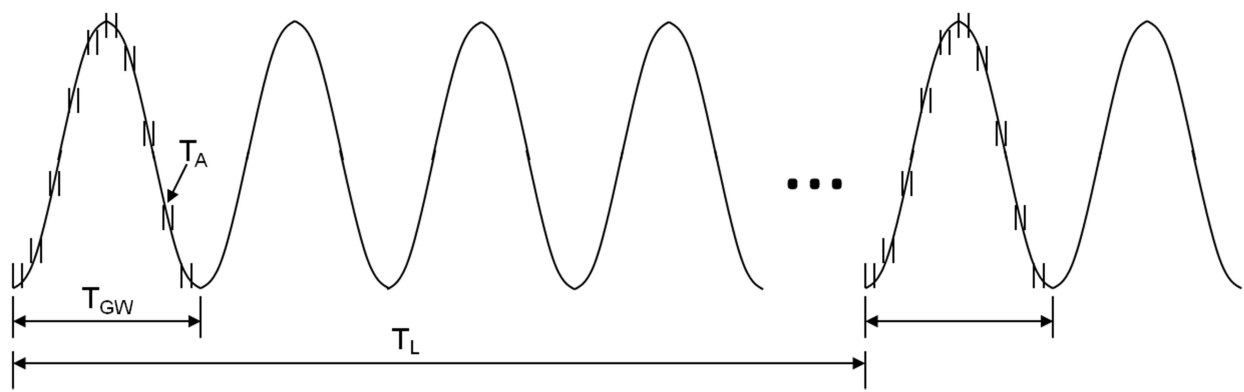

Fig. 8 Time relationship of the arm sampling and the loop operation